\documentclass[journal,twoside,web]{ieeecolor}
\usepackage{jsen}
\usepackage{cite}
\usepackage{amsmath,amssymb,amsfonts}
\usepackage{algorithmic}
\usepackage{graphicx}
\usepackage{textcomp}
\usepackage{wrapfig}
\usepackage{tabularx}
\usepackage{stfloats}
\usepackage{url}
\usepackage{verbatim}
\usepackage{subcaption}
\usepackage{multirow}
\usepackage{makecell}
\let\labelindent\relax
\usepackage{enumitem}%
\usepackage[utf8]{inputenc}

\DeclareUnicodeCharacter{2212}{-}

\def\BibTeX{{\rm B\kern-.05em{\sc i\kern-.025em b}\kern-.08em
    T\kern-.1667em\lower.7ex\hbox{E}\kern-.125emX}}
\definecolor{abstractbg}{rgb}{255,255,255}
\begin{document}

\title{GreenScan: Towards large-scale terrestrial monitoring the health of urban trees using mobile sensing}

\author{
    \authorblockN{Akshit Gupta\authorrefmark{1}\authorrefmark{2}, Simone Mora\authorrefmark{2}, Fan Zhang\authorrefmark{3}\authorrefmark{2}, Martine Rutten\authorrefmark{1}, R. Venkatesha Prasad\authorrefmark{1}, Carlo Ratti\authorrefmark{2}}
    \\
    \authorblockA{\authorrefmark{1}Delft University of Technology
    }\\
    \authorblockA{\authorrefmark{2}Massachusetts Institute of Technology 
    }\\
    \authorblockA{\authorrefmark{3}Peking University 
    }
\thanks{This work was supported by the MIT Senseable City Lab Consortium. A. Gupta would like to thank Renswoude Foundation, FAST Delft and EFL Stichting for their financial support.}
\thanks{Akshit Gupta is the corresponding author. He was a visiting student researcher at the MIT Senseable City Lab at the time of this work, while studying at TU Delft (email: a.gupta-5@tudelft.nl)}
}


\IEEEtitleabstractindextext{%
\fcolorbox{abstractbg}{abstractbg}{%
\begin{minipage}{\textwidth}%

\begin{abstract}
Healthy urban greenery is a fundamental asset to mitigate climate change phenomena such as extreme heat and air pollution. However, urban trees are often affected by abiotic and biotic stressors that hamper their functionality, and whenever not timely managed, even their survival. While the current greenery inspection techniques can help in taking effective measures, they often require a high amount of human labor, making frequent assessments infeasible at city-wide scales. In this paper, we present GreenScan, a ground-based sensing system designed to provide health assessments of urban trees at high spatio-temporal resolutions, with low costs. The system utilises thermal and multi-spectral imaging sensors fused using a custom computer vision model in order to estimate two tree health indexes. The evaluation of the system was performed through data collection experiments in Cambridge, USA. Overall, this work illustrates a novel approach for autonomous mobile ground-based tree health monitoring on city-wide scales at high temporal resolutions with low-costs.
\end{abstract}

\end{minipage}}}

\maketitle

\section{Introduction}
\label{sec:introduction}

\IEEEPARstart{U}{rban} greenery improves the resilience of cities to climate change. Nowadays, protecting, managing, and restoring greenery ecosystems is fundamental for climate-resilient development, given the multiple risks posed to humanity and nature by global warming and climate change as per the latest UN-IPCC Report \cite{ipcc}. In cities,  tree canopies and vegetation provide a wide range of ecosystem services such as air filtering, carbon sequestration, reduced energy consumption, increased biodiversity and decreased local temperatures \cite{GREGORYMCPHERSON199241,Hobbie_2020}. However, urban trees are experiencing an ample amount of \textit{abiotic} stressors (e.g. soil salinity, heat waves) and \textit{biotic} stressors (caused by living agents such as insects and bacteria) that are exacerbated due to climate change \cite{climateChangeAndGreenery1, climateChangeAndGreenery2,allan2021ipcc}. As a result, their functionality, productivity, and survival are of increasing concern \cite{climateChangeGreenery3}. Trees with poor health cannot provide most of their beneficial ecosystem services \cite{urbanTreeManagementEcosystemDelivery, tree_mortality}. For instance, trees with low transpiration rates do not cool the environment sufficiently and trees with low growth rates have a reduced shading effect. By 2050, it is expected that about two-thirds of urban tree species worldwide will fail to provide the desired climate-positive benefits \cite{natureClimateChange}.

The practice of measuring and monitoring urban trees began over a century ago \cite{solotaroff1912shade}. Today, the health of trees can be monitored using manual inspection by arborists with good quality results \cite{classificationtreedecaydetection}; yet the high labor cost lead leads to assessments performed infrequently at very low temporal resolutions such as once every 3-5 years. Technology-assisted monitoring methods can complement manual inspections \cite{Gupta:2024aa}. However, these methods are impeded by variable data quality, low spatial granularity (remote sensing), or high operational costs (airborne sensing) \cite{sensors-21-00295-v22}. Further, most of these methods are unable to quantify the vegetation elements below the tree canopy such as green walls, short trees, or shrubs \cite{gsvImagesGreeneryExtraction,CUFMPAppendix}. All these challenges lead to the lack of urban tree health data in cities and appropriate urban forest management. For instance, adverse health conditions in trees being discovered only after severe damage is already inflicted. Further,  from an urban planning perspective, intricate relationships of urban trees with other micro-scale ecosystem services such as air quality improvements and benefits to public health are difficult to quantify. For instance, inappropriate placement of trees in outdoor environments can be detrimental as they can serve to trap air pollutants \cite{greeneryMitigationNatureReview}.



Recently, several projects have investigated developing novel alternatives for environmental sensing. For instance, applying AI (Artificial Intelligence) based methods on GSV (Google Street View) images to detect the presence of trees \cite{treepedia1, treepedia2}, or using drive-by strategies to measure air pollution \cite{IEEECityScanner, 8767186} in a cost efficient way. For instance, it was demonstrated that just ten random taxis could capture data over one-third of streets in Manhattan (New York City) in a single day using drive-by sensing \cite{urbanSensing}. Additionally, citizen science-based approaches \cite{SILVERTOWN2009467} have also been successful in measuring urban environmental parameters \cite{eubook}. All these methods are set within the domain of opportunistic sensing and are aimed at developing platforms that can be deployed and operated without the need of an expensive or a dedicated infrastructure. Thus, allowing democratic access to even under-resourced cities that are affected by climate change in a disproportionate manner \cite{cd04898}.




Following on this trend and the critical need for protecting and managing urban forestry, in this work, we develop a novel system, named GreenScan, which measures the health of urban trees on city-wide scales from ground level (terrestrially). The system fuses high-quality data from low-cost thermal and multispectral imaging sensors using custom computer vision models to generate two complementary tree health indexes namely NDVI (Normalised Difference Vegetation Index) and CTD (Canopy Temperature Depression) which respectively indicate the photosynthetic capacity and water stress levels of a tree. GreenScan was designed both for deployment in citizen-science paradigms by being carried by pedestrians, or in drive-by sensing approaches by being mounted on urban vehicles such as taxis and garbage trucks. Thus, enabling terrestrial urban tree health measurements with high spatial and temporal resolutions at low costs for cities and municipalities around the world.

In this paper, we first give a brief overview of the state-of-the-art technology tools and methods to monitor the health of urban greenery with focus on low-cost solutions. We present the design of GreenScan and describe the implementation of the hardware and software components. We evaluate the system with forty urban trees in uncontrolled outdoor environments and analyse the performance of the system. Finally, we conclude by identifying the immediate future research that can be enabled through large scale deployment of GreenScan while discerning the limitations of this work.

\section{Related Work}

Currently, the health of trees is monitored through manual inspection by human experts, remote and airborne sensing through satellites or UAVs, direct installation of embedded sensors on/near the tree, handheld imaging based sensing or opportunistic sensing using street view imaging \cite{Gupta:2024aa}. A comparison of these methods in terms of working mechanism, cost, and quality of assessment is shown concisely in Table \ref{comparison}.

\begin{table*}[tb]
\centering
\caption{A comparison of sensing approaches along with the working mechanism, cost, and quality of assessment to analyse tree health \\ \textit{(\mbox{*} refers to relative cost where \$ is the lowest cost and \$\$\$\$ is the highest cost for large-scale evaluation of multiple trees based on the scale in \cite{classificationtreedecaydetection})}}
{\begin{tabularx}{\linewidth}{|X|X|X|l|}
\hline
\textbf{Approach}   & \textbf{Principle}  & \textbf{Quality of Assessment}        & \textbf{Cost\mbox{*}} \\ \hline
\textbf{Manual inspection} & Depends on method   & Generally high, varying based on method        & \$\$\$\$ \\ \hline
\textbf{Embedded sensing} & Depends on method     & High but lower than manual inspection               & \$\$\$ \\ \hline
\multirow{3}{*}{
\begin{tabularx}{1.04\linewidth}{X|X}
\multirow{3}{*}{\textbf{\shortstack{Handheld\\imaging}}}    & \textbf{Multi/Hyper-spectral imaging} \\
\cline{2-2}
                         & \textbf{\makecell[l]{Thermal \\ \hphantom{2} }} \\
                         \cline{2-2}
                         & \textbf{LiDAR}
    
\end{tabularx}
}
& Properties of chlorophyll (photosynthesis) and cell structure  & High-quality quantitative value  & \$\$       \\ \cline{2-4}
& Cavities, temperature gradient and water stress & Cavities: qualitative value, water stress and temperature gradient: quantitative value & \$ \\ \cline{2-4}
& Geometrical parameters such as Leaf Area Index (LAI) and leaf density & Low-quality quantitative value & \$ to \$\$ \\ \hline
\textbf{Street-view based (visible spectrum)}   & Uses image processing to quantify amount of greenery   & No health assessment, only quantity of greenery & \$ \\ \hline
\textbf{Remote sensing (multi/hyper-spectral, thermal imaging, LiDAR)}  & Depends on the type of imaging  & Top-level view only & \$ \\ \hline
\end{tabularx}}
\label{comparison}
\end{table*}

Manual inspection involves the work of arborists (human-experts) inspecting trees visually, often with the aid of tools such as borers (to extract a wooden core sample from the tree for laboratory analysis) or resistographs (to measure the electrical resistance of the trunk). These methods usually provide a high-quality assessment, but they are time-consuming due to the amount of human labor involved to perform a tree-by-tree assessment. Further, although effective, methods that require drilling and penetration in the living wood may create an entry path for pathogens or may alter the structural integrity of a tree. For a review on these methods, see \cite{classificationtreedecaydetection}. 

Embedded sensing involves the deployment of sensors in the bark of a tree (the outer wooden part of a tree) or in the soil. These sensors can rely on physical, chemical or electrical phenomenon to detect the presence of parasites, e.g. detecting sudden minimal bark vibrations produced by parasites locomotion and feeding \cite{staticSensorInsectDetection} as well as water uptake and transpiration, e.g. measuring electrical impedance using a pair of electrodes placed in the trunk at opposite positions \cite{staticElectricalImpedanceSpectroscopyInVivo}. These methods generate data at high temporal resolutions with little or no human supervision required; yet at the cost of installing and maintaining one or more sensors per tree. For a review on these methods, see \cite{BoverallGeneralReviewOfMonitoring}.

Imaging-based methods involve the use of optical sensors such as thermal imaging sensors, HMI (hyperspectral or multispectral imaging) sensors or LiDAR (Light Detection and Ranging). Thermal imaging is based on IR (InfraRed) radiation emitted from materials and it is mainly used to (i) measure cavities and physical damages in the living wood \cite{OverviewofThermalImagingforTreeAssessment,THermalImagingTreeHealthDetection}, (ii) detect infections caused by insects and bacteria \cite{uavThermalCamera,ContradictionthermalCameraTempRiseDisease}, and (iii) calculate water stress levels by measuring the temperature of the leaves in the canopy \cite{sensors-21-00295-v22}. On the other hand, HMI sensors capture various bands in the electromagnetic spectrum, usually near-infrared and parts of the visible spectrum. This captured data is utilised to calculate various vegetation indexes, the most popular being NDVI (Normalized Difference Vegetation Index). 
HMI sensors are often used for remote sensing applications \cite{renewRemoteSesningForestHeath}, although static sensors also exist \cite{sensors-20-03208-v2}. Calibration methods are critical to achieve quality results \cite{Huang2021}. LiDAR (Light Detection and Ranging) sensors, can be utilised to measure geometrical parameters such as the leaves surrounding a branch, trunk diameter, etc. and to estimate the LAI (Leaf Area Index) \cite{treeHealthAndLidar}. However, contradictory studies have been observed on the usage of LiDAR with some works such as \cite{mappingTreeHealthLidarLAIInLab} claiming no increase in health classification performance with its addition.
Usually, LiDAR and HMI sensor approaches are often deployed in tandem in both airborne \cite{treeHealthAndLidar} and ground-based \cite{combineALSMLSLidarThermalforTreehealthTreeDetectionAndSpeciesClassification} approaches. 


Recently, street-view based methods based on RGB (Red, Green, Blue) images usually involving the use of google-street view images have became popular. These methods are utilised either to quantify the presence of urban greenery \cite{treepedia1,BiaduStreetViewGreeneryIndexDL}, catalog species \cite{DBLP:journals/corr/abs-1910-02675}, and shading effects \cite{gsvShadingTreesCalculation}. While these approaches are cost-effective and scalable, they are only able to quantify the extent of urban greenery at a terrestrial level rather than its health.

When imaging-based sensors are deployed on satellites, airplanes or UAVs (Unmanned Aerial Vehicles), high spatial coverage can be achieved. However, satellites have a low temporal resolution due to infrequent re-visit time and data quality being dependent on the availability of clear skies \cite{sensors-21-00295-v22}. Data collection using UAVs and airplanes involves high operational costs and is unsuitable for highly urbanised environments due to aviation regulations. Most importantly, both airborne sensing and satellite imagery can only capture an overhead view of urban tree canopies. As a result, lower vegetation elements such as green walls, short trees, or shrubs are often missed or misinterpreted \cite{gsvImagesGreeneryExtraction}.

For a systematic review of the technological methods and tools for greenery health monitoring, see \cite{Gupta:2024aa}.

\subsection{Research gaps and influence on design}
\label{sec:ResearchGap}


We seek to provide a scalable system that provides high quality data with low-costs. 
Comparing the different approaches in Table \ref{comparison}, it emerges that ground-based (terrestrial) sensing approaches combined with imaging-based methods can look at vegetation elements in a holistic manner with high quality data gathered either through drive-by sensing or citizen science paradigms. Additionally, the advances in deep learning based computer vision models for imaging data in the past decade enables the development of a system that is broad in scope. Narrowing this down, the past studies measuring tree health from ground level and utilising low-cost imaging sensors are also shown concisely in Table \ref{tab:comparisonotherWorks}. These studies are limited by requiring manual analysis of images by humans \cite{AUsesThermalCameraIRThermometerCheaperNDVISensorMultispectralCamera}, outputting only raw data without ground truth validation \cite{sensors-21-00295-v22, ApplyingMLOnThermalImagesClusteringManualDataset} or requiring controlled system deployment and operation \cite{sensors-21-00295-v22,thermalWaterStress,robotLIDATNVDISensor}. 

Our work builds upon all these insights. Hence, in GreenScan, we utilise HMI imaging and thermal imaging sensors to autonomously measure two health indexes namely NDVI and CTD from ground level. GreenScan is designed to be completely autonomous (by utilising computer vision model based on deep learning), is suitable to be deployed in non-controlled environments and the early results are compared with a ground truth dataset provided by a municipality. While data from HMI and thermal imaging can be used to generate a number of health indices, we carefully chose NDVI and CTD indices after explicit considerations. Our choices are driven by 1) NDVI remains one of the most important and popular indices used in the domain \cite{Huang2021} and the ground truth dataset provided by the municipality contains remote NDVI for a fair scientific comparison 2) CTD is one of the relatively simpler metrics for assessing properties of a tree such as water consumption and its resilience to drought and heat stress events \cite{Lepekhov2022-vc}, and it uses different wavelengths than NDVI (thermal imaging sensors instead of HMI sensors). In turn, generating two complementary health parameters for urban trees.

\begin{table}[h]
\caption{A concise comparison of our work with earlier works in the field measuring tree health terrestrially}
\begin{tabularx}{\linewidth}{|l|X|X|X|X|}
\hline
\textbf{Works}                       & \textbf{\begin{tabular}[c]{@{}l@{}}Autonomous\\ (No human \\ intervention\\ needed)\end{tabular}} & \textbf{Approach}          & \textbf{\begin{tabular}[c]{@{}l@{}}Ground \\ Truth \\ Comparison\end{tabular}} & \textbf{Evaluation}                   \\ \hline
\cite{sensors-21-00295-v22}                            & Yes                                                                                               & Mobile (Cars)              & No                                                                             & 172 trees (only raw system output with no comparison)   \\ \hline
\cite{Ballester2013UsefulnessOT} and \cite{thermalWaterStress} & Yes                                                                                               & Handheld                   & Yes                                                                            & 44 images (trees not mentioned)       \\ \hline
\cite{ApplyingMLOnThermalImagesClusteringManualDataset}                         & Yes                                                                                               & Handheld                   & No                                                                             & 8 trees                               \\ \hline
\cite{53359897} and \cite{robotLIDATNVDISensor}                      & Yes                                                                                               & Mobile (Robot)             & Yes                                                                            & 2 trees (in controlled lab environment) \\ \hline
\cite{AUsesThermalCameraIRThermometerCheaperNDVISensorMultispectralCamera}                   & No                                                                                                & Mobile (Cars)              & Yes                                                                            & 20 trees                              \\ \hline
\textbf{This work}                            & \textbf{Yes}                                                                                               & \textbf{Mobile (Cars) and Citizen Science} & \textbf{Yes}                                                                            & 40 trees                              \\ \hline
\end{tabularx}
\label{tab:comparisonotherWorks}
\end{table}

\section{Methodology}

\begin{figure}[t]
\begin{subfigure}[t]{\linewidth}
    \centering
    \includegraphics[width=0.5\linewidth]{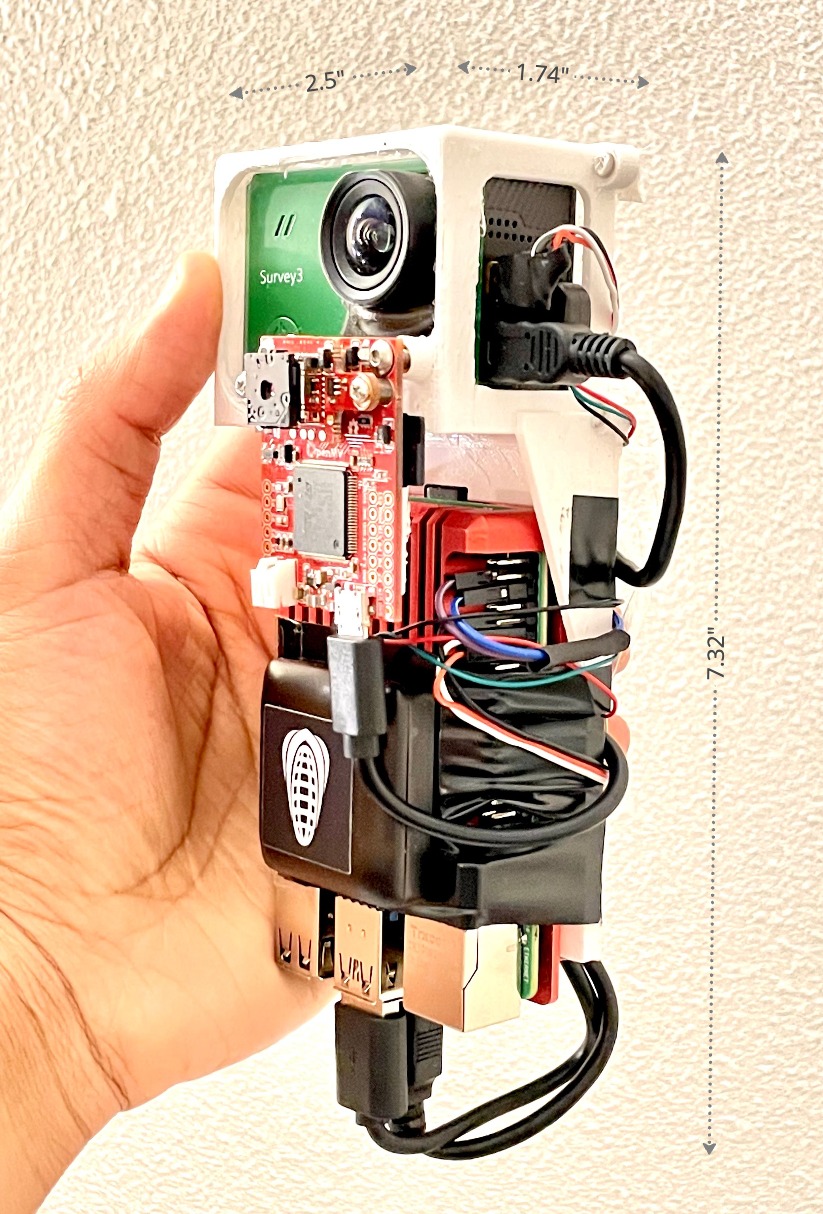}
    \caption{ All hardware components encased with the 3D printed case. Dimensions in inches: 7.32” X 2.50” X 1.74”}
    \label{fig:casing}
\end{subfigure}
\centering
\begin{subfigure}[t]{0.4\linewidth}
    \centering
    \includegraphics[width=\linewidth]{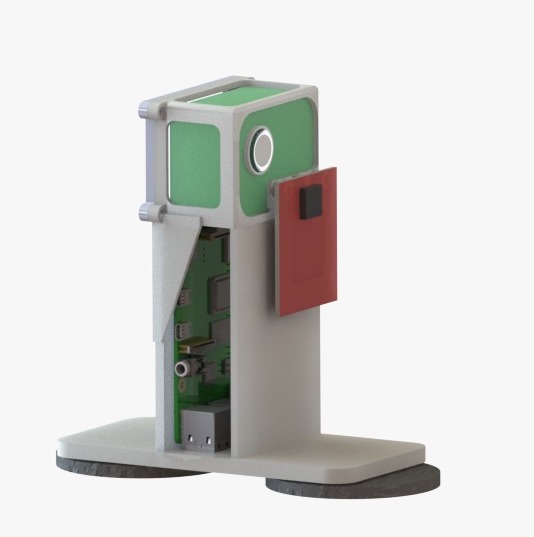}
    \caption{Concept casing with magnets}
    \label{fig:concept1} 
\end{subfigure}
\hfil
\begin{subfigure}[t]{0.4\linewidth}
    \centering
    \includegraphics[width=\linewidth]{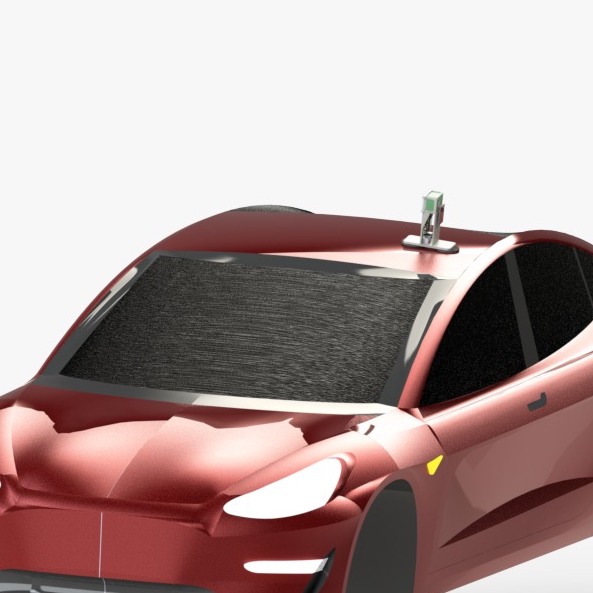} 
    \caption{The system attached on the top of a car}
    \label{fig:concept2}
\end{subfigure}
\begin{subfigure}[t]{0.4\linewidth}
    \centering
    \includegraphics[width=\linewidth]{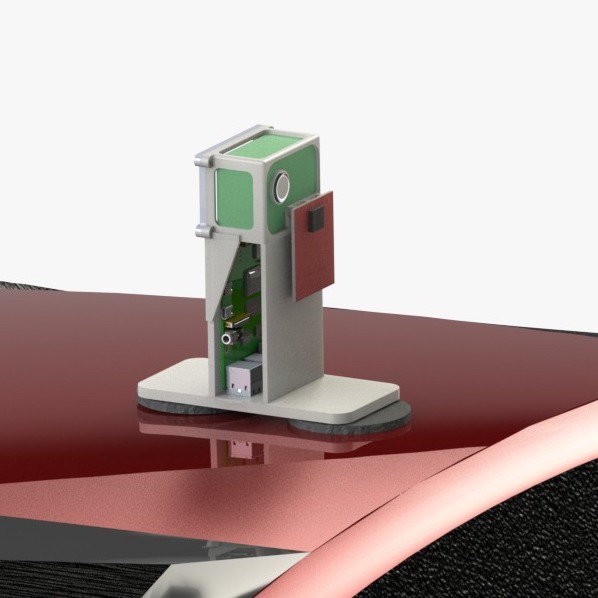} 
    \caption{Closeup view of the system on the roof of a car}
    \label{fig:concept3}
\end{subfigure} 
\caption{A visualization of the current GreenScan system and the concept casing}
\end{figure}

The GreenScan system integrates low-cost thermal and multispectral imaging sensors which are attached to a single board computer. The system processes the imaging data generated by these sensors using a custom computer vision model to generate the two tree health indexes namely CTD and NDVI. All these components were encased in a 3D printed case as shown in Figure \ref{fig:casing}. The case was designed such that it is suitable to be attachable to moving vehicles without any alterations using magnets, as shown in Figures \ref{fig:concept1}, \ref{fig:concept2} and \ref{fig:concept3}. In this section, we aim to explain all the major modules of GreenScan.

\subsection{System Architecture}
\label{sec:architecture}

\begin{figure*}[ht]
    \centering
    \includegraphics[width=0.75\linewidth]{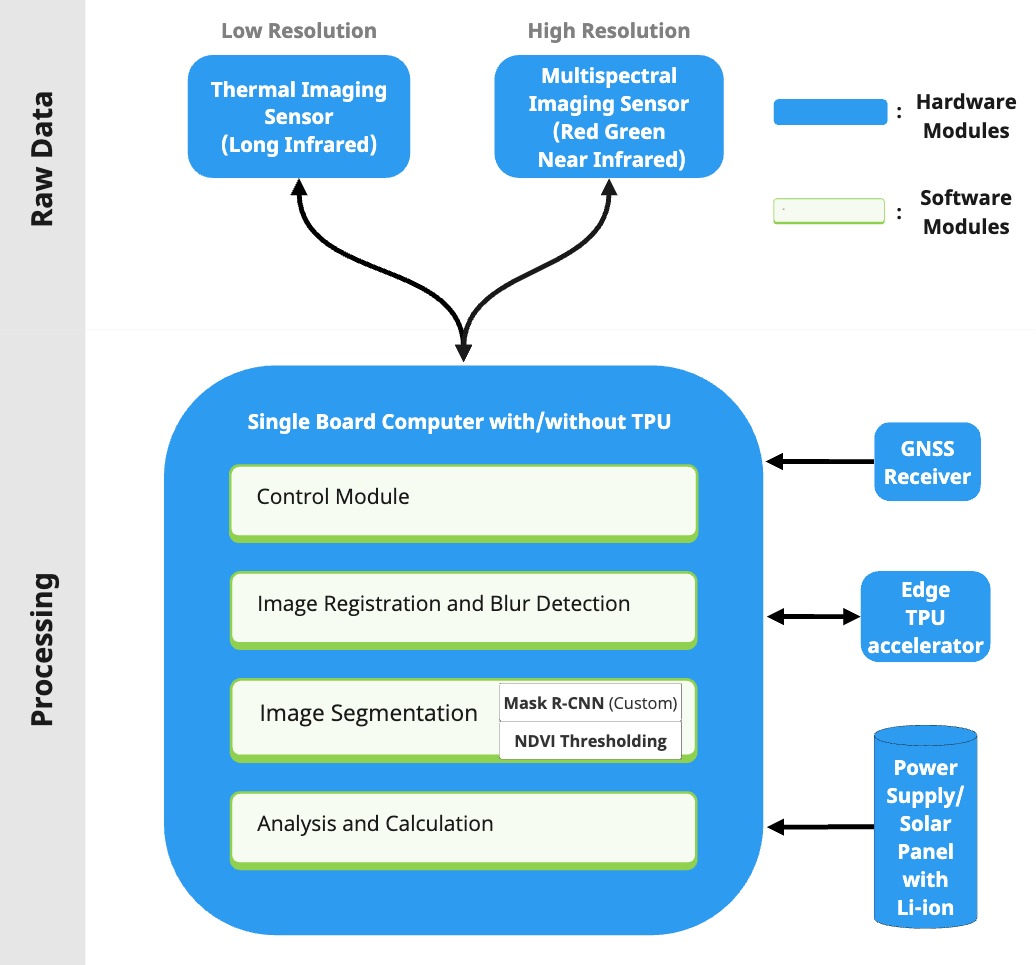}
    \caption{Architecture Diagram of the GreenScan system}
    \label{fig:architecture}
\end{figure*}

The block diagram of the entire GreenScan system architecture is shown in Figure \ref{fig:architecture}. The first five modules are related to hardware, while the remaining four modules are related to software.


\subsubsection{Hardware Modules}
In the following, we first provide the generic description of each hardware module followed by the concrete implementation of the same in the GreenScan system.

\begin{description} [style=unboxed,leftmargin=0cm]
    \item \textbf{1. Thermal Imaging Sensor}: A thermal imaging sensor with radiometric calibration (to measure true temperature) is attached to the central single board computer and captures long-wave infrared images normalised to a suitable temperature range with low pixel resolution. A narrow temperature range is preferred to decrease the effect of non-linear noise across the sensor as the low cost thermal imaging sensors are constrained in terms of resolution. This long-wave infrared imaging data is used for generation of CTD (Canopy Temperature Depression) which indicates the water stress levels of a tree. 
    
     For concrete implementation, we used FLIR Lepton 3.5 (spectrum: longwave-infrared @8 \textmu m - 14 \textmu m) attached to an OpenMV cam H7 using a FLIR Lepton adapter module. This captured thermal images with a pixel resolution of $160\times120$ which are normalised to a suitable temperature range ($ -10^\circ $ to $ 40 ^\circ $ C). This temperature range was chosen based on the lowest and highest temperature ($ \pm  10 ^\circ $C) of trees found during the data collection experiments (in Section \ref{sec:evaluation}). The OpenMV cam H7 communicates with Raspberry Pi via RPC (remote procedure call) over USB.

    \item \textbf{2. Multispectral imaging sensor}: A multispectral imaging sensor is attached to the central single board computer and captures RGN (Red, Green and Near-infrared) imaging data with high pixel resolution. The near-infrared and red imaging data is used for generation of the NDVI (Normalised Difference Vegetation Index). Further, this high resolution imaging data is used for segmentation of the tree canopy from the images using the custom deep learning model as described in the \textit{Image Segmentation Module}.
    
    For implementation, MAPIR Survey 3W (spectrum:  red@660nm, green@550nm, near-infrared@850nm) was attached to the Raspberry Pi over USB and captured RGN imaging data with a pixel resolution of $4000\times3000$. To control the MAPIR Survey 3W for triggering the capture and transfer of images, PWM (pulse width modulation) signals over the micro-HDMI port of MAPIR Survey 3W are utilised.
    
    \item \textbf{3. GNSS Receiver}: A GNSS (Global Navigation Satellite System) receiver with support for GPS, GLONASS, and Galileo is used to find the current location of the system and geo-tag all the images of the trees captured. 
    
    For this prototype, the RGN images captured were geo-tagged using the standard GPS adapter available for MAPIR Survey 3W.

    \item \textbf{4. Single Board Computer with/without Edge TPU}: A single board computer without/with onboard edge TPU (Tensor Processing Unit) or USB edge TPU accelerator (such as USB TPU accelerator from coral.ai) acts as the central brain of the system integrating all the hardware and software components. The edge TPU allows to speed up deep learning operations, improving the images processed per second without sending any data to the cloud.
    
    Raspberry Pi 3 was utilised as the single board computer in the GreenScan system running the software modules (see Section \ref{subsec:software}).

    \item \textbf{5. Power Supply / Solar Panel}: A lithium-ion battery (10000 mAh) is used to ensure uninterrupted power supply to the system along with support for charging over a solar panel or a standard power adapter (5V/2A).
\end{description}

\subsubsection{Software Modules}
\label{subsec:software}
\begin{figure}[ht]
    \centering
    \includegraphics[width=1\linewidth]{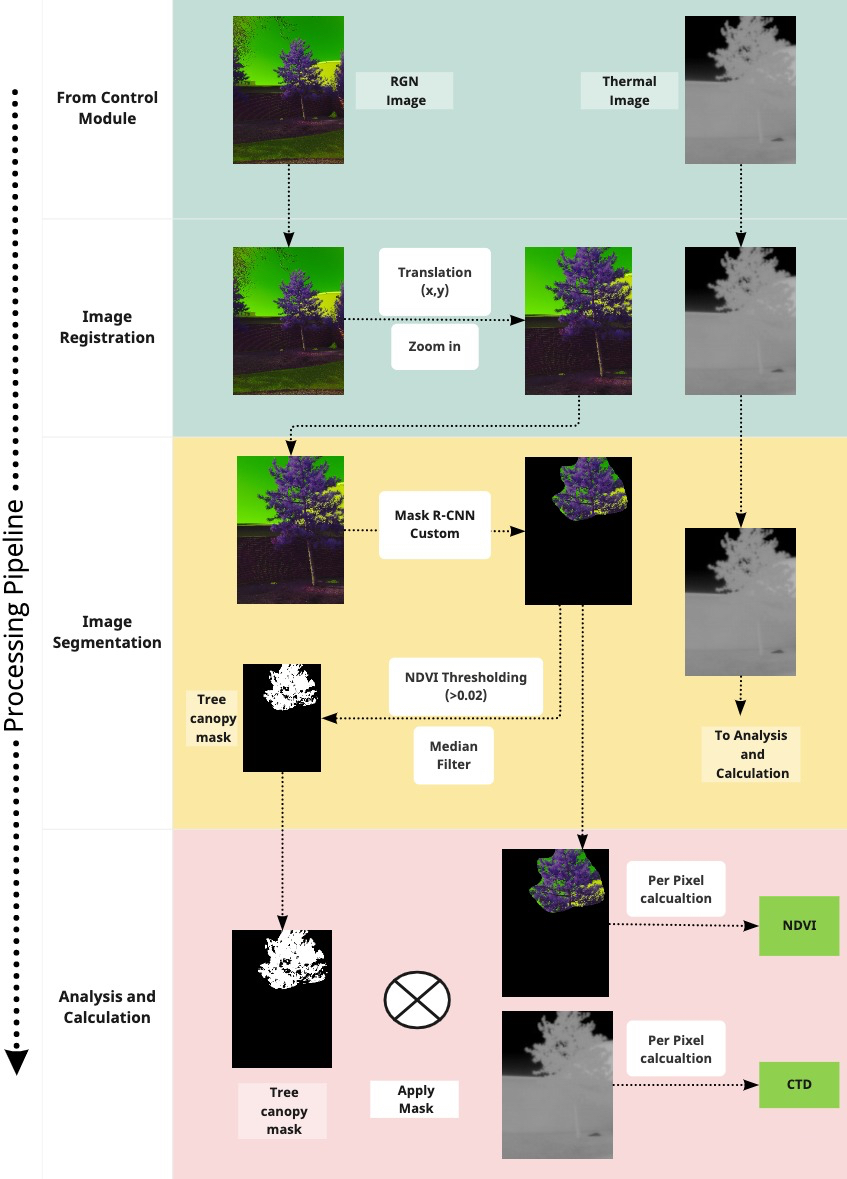}
    \caption{A visualisation of processing the images at each software module on the Raspberry Pi}
    \label{fig:imageProcessingPipeline}
\end{figure}

Herewith, we provide the description of each software module followed by the concrete implementation of the same in the GreenScan system. A visualization of processing the images after each software module is also shown in Figure \ref{fig:imageProcessingPipeline}.

\begin{description}[style=unboxed,leftmargin=0cm]
    \item \textbf{1. Control Module}: This module handles all the embedded communication with the hardware. It includes detecting the event trigger, signaling the sensors to capture the images, and transferring the captured images to the central single board computer.
    
    The event trigger signals the beginning of processing on the Raspberry Pi and in the current prototype, a press of a push button is used as an event trigger for the data collection experiments. This event trigger can also be the co-location of the system with particular GPS coordinates fetched from a tree inventory database. For the thermal imaging sensor, this involves the initiation of callbacks requesting the transfer of the current image frame from FLIR Lepton 3.5. For the multispectral imaging sensor, this involves generating PWM signals to capture an image, mounting the memory card installed in the MAPIR Survey 3W with the Raspberry Pi, transferring the captured image to Raspberry Pi, and finally, unmounting the memory card from the Raspberry Pi.
    
    \item \textbf{2. Image Registration}: Image registration involves matching or aligning images taken by two different sensors into a single coordinate system for further analysis \cite{imageRegistration}. It includes detecting key points from one image and mapping them to another image. Since both the multispectral and thermal imaging sensors have different FoVs (field of views) and are un-aligned, this modules aligns the multispectral images to the thermal images through linear translation in both horizontal and vertical directions. Also, to compensate for wider FOV of the multispectral sensor, this module also handles zooming in on the multispectral images. 
    
    For the current prototype, the values of translation in X and Y directions were found to be +50 (right) and +150 (upwards) pixels respectively and the zoom scale was found to be 0.57 (where 1 indicates no magnification and 0 indicates 100 \% magnification) to perfectly overlay the thermal and RGN images. These parameters were found by manually taking multiple RGN and thermal images and overlaying them. An instance of inputs and outputs utilizing this module are shown in Figure \ref{fig:imageProcessingPipeline}. 
    
    Further, automatic image registration using three image registration algorithms namely SIFT, SURF, and ORB \cite{autoimageRegister} was also tested. However, these algorithms were not able to detect useful keypoints or features in the thermal images possibly due to the low resolution (160x120).
    
    \item \textbf{3. Image Segmentation}: The is the most computationally intensive software module of the system. Recall that the aim of our system is to calculate the NDVI and CTD values for each tree in the images. However, these values should be calculated only for the leaves in tree canopy excluding the wooden parts, such as the trunk and branches. This is solved using a fusion of custom developed Mask R-CNN (Mask Regional Convolutional Neural Network) and pixel-wise NDVI analysis. Mask R-CNN \cite{maskrcnn} is an object detection and instance segmentation model that identifies and then draws a precise mask around the detected object. Given a multispectral RGN image captured using the multispectral imaging sensor, this task can be broken into two sub-problems as follows:
    
    \begin{itemize}
        \item \textbf{Detect the canopy part of the trees even in cases where the image contains multiple trees}:
        This is solved using a custom-developed Mask R-CNN model. The Mask R-CNN model is trained using transfer learning and discussed in more detail in Section \ref{sec:customMaskRCNN}. It segments the instances of the tree canopies in the RGN image by generating a mask (segmentation) over them as shown in Figure~\ref{fig:maskrcnnOutput}.
        
        \item \textbf{Remove Noise}:
        Once the canopy of the tree is detected, the segmentation of only the leaves of the tree without the wooden branches and sky. The non-vegetation elements such as trunks, branches, and sky have very low NDVI values compared to vegetation elements which have significantly higher NDVI. Thus, we employ a thresholding based method which first calculates the individual NDVI of each pixel in the segmentation mask generated by Mask R-CNN and then eliminates pixels with NDVI values below a certain threshold. The calculation of NDVI for each pixel is computed by plugging the raw values of the red and near-infrared channels of the pixel in (\ref{eq:NDVI}). In order to eliminate noise along the edges of tree canopy, median filtering is also employed. 
    \end{itemize}
    The end result employing the above two-stage approach gives segmentation of only leaves present in the tree canopy while eliminating the sky, wooden branches, trunk and other street objects such as buildings and cars in the multispectral image. Since both the thermal and multispectral images are registered, the same mask of leaves in tree canopy can also be used for thermal images.  
    
    With the MAPIR Survey 3W employed in the GreenScan system, a value of 0.02 was used as the cutoff to eliminate non-vegetation elements in the image. This value was derived using the analysis of the images captured during data collection experiments. An instance of inputs and outputs utilizing this module is also shown in Figure \ref{fig:imageProcessingPipeline}.
    
    \item \textbf{4. Analysis and Calculation Module}: This module handles the calculation of final NDVI and CTD for each tree in the field of view of the imaging sensors.
    
    The CTD value is computed by calculating the raw temperature value for each pixel in the grayscale thermal image as per (\ref{eq:tempPixel}), computing the mean temperature over all pixels in the canopy and subtracting the ambient air temperature from the mean canopy temperature as per (\ref{eq:CTDcalulation}). 
   CTD is calculated as:
    \begin{equation}
    CTD = T_{canopy}-T_{air}
    \label{eq:CTD}
    \end{equation}
    where $T_{canopy}$ and $T_{air}$ are canopy temperature and air temperature respectively in $^\circ $ C. 
    
    The temperature of each pixel is calculated as:
    \begin{equation}
    T_{pixel} = \frac{P_{value}}{255} * (T_{max} - T_{min}) + T_{min}
    \label{eq:tempPixel}
    \end{equation}
    where $P_{value}$ is the pixel value in normalised thermal image, $T_{min}$ and  $T_{max}$ are configured temperature range for the thermal imaging sensor respectively ($ -10^\circ $ C and $ 40 ^\circ $ C in our case).
    Then, as per (\ref{eq:CTD}), CTD is calculated as: 
    \begin{equation}
    CTD = \overline{T_{pixel}}-T_{air}
    \label{eq:CTDcalulation}
    \end{equation}
    where $\overline{T_{pixel}}$ is the average canopy temperature for all segmented pixels in the image and $T_{air}$ is the air temperature respectively.
    
    To calculate the NDVI, each pixel in the RGN image is split into its three constituting channels (red, green and near-infrared). The raw NDVI value for each pixel is calculated from red and near-infrared channels as per (\ref{eq:NDVI}). To compensate for the aperture adjustment, the focal adjustment and other mechanical adjustments performed by the multispectral imaging sensor, the raw NDVI is normalised by applying a correction factor similar to the dynamic range of a camera \cite{dynamicrange} as shown in (\ref{eq:correctedNDVI}).
    NDVI is calculated as:
    \begin{equation}
    NDVI = \frac{NIR-Red}{NIR+Red},
    \label{eq:NDVI}
    \end{equation}
    where $NIR$ and $Red$ are values of near-infrared channel and visible red channel for each pixel respectively.
    The corrected NDVI is calculated as:
    \begin{equation}
    NDVI_{corrected} = \frac{NDVI_{raw}}{|NDVI_{max}|} * |NDVI_{min}|
    \label{eq:correctedNDVI}
    \end{equation}
    where $NDVI_{raw}$ is the raw NDVI of a pixel, $NDVI_{max}$ and  $NDVI_{min}$ are maximum and minimum NDVI values for all pixels in the segmented image.
    
Finally, the NDVI for the entire canopy is computed by taking the mean over the corrected NDVI values for all pixels consisting of leaves in the segmented tree canopy.
\end{description}

\subsection{Development of Custom Mask R-CNN}
\label{sec:customMaskRCNN}
 For the system to operate autonomously, the images will be captured in an unsupervised fashion. Thus, in addition to multiple trees in a single image, they may contain other objects such as cars, buildings, grass, and snow. Hence, it is imperative to individually identify all the tree canopies in an image and feed them to the \textbf{Analysis and Calculation Module}. The custom mask R-CNN part of the image segmentation module solves this by providing instance segmentation of the tree canopies in the image. To our knowledge, there is no pre-existing model available for instance segmentation of tree canopies or even trees in standard RGB images. The problem is further complicated as our input is RGN (Red, Green, Near-infrared) images from the multispectral imaging sensor instead of standard RGB images. For instance, we found that pre-trained models like Deeplabv3 \cite{deeplabv3}, which can perform semantic segmentation of trees and vegetation on standard RGB images, perform poorly on RGN images.

\subsubsection{Training Data}
\label{sec:datasetCreation}
Any deep learning model requires training data in order to optimise the weights and activations of the layers. However, there does not exist a dataset with labels for instances of trees or tree canopies for RGN Images. Hence, we manually created the dataset using the RGN images collected during the data collection experiments (See Section \ref{subsec:dataCollection}). Each tree canopy in the image was manually annotated using a popular image annotation tool called LabelMe \cite{labelme}. During annotation, only tree canopies that were completely present in the image were labelled. After this process, our dataset consisted of 51 annotated RGN images with two classes namely tree canopies and background. 

\subsubsection{Training Process and Training Curve}
\label{subsec:ModelTraining}

\begin{figure}[ht]
    \centering
    \includegraphics[width=0.9\linewidth]{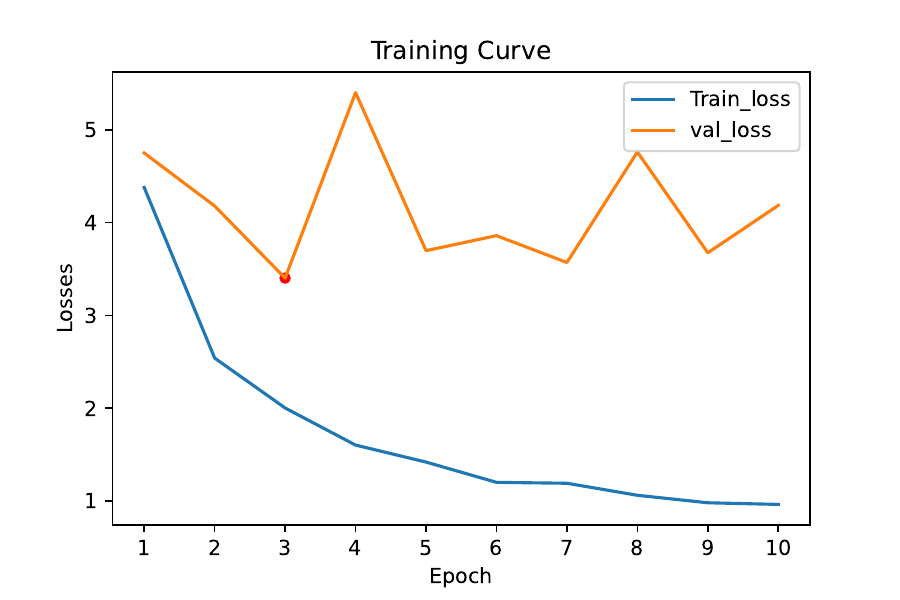}
    \caption{The training curve of Mask R-CNN with epochs=10 and batch size=4. The red point indicates point of minimum loss and training losses are as defined in \cite{maskrcnn}}
    \label{fig:trainingCurve}
\end{figure}

\begin{figure}[t] 
\centering
  \begin{subfigure}[t]{0.45\linewidth}
    \centering
    \includegraphics[width=0.9\linewidth]{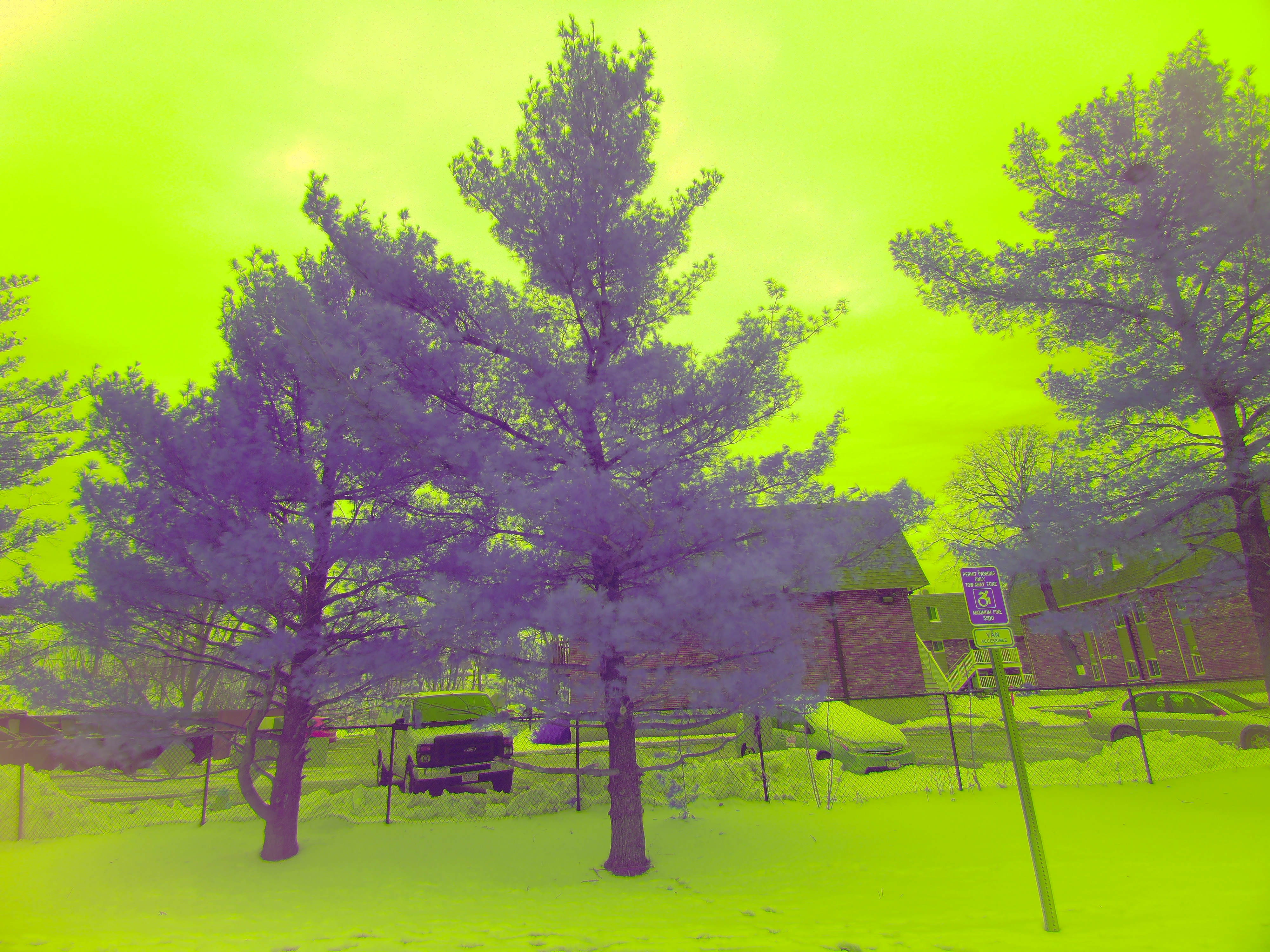} 
    \caption{Input RGN Image captured from MAPIR Survey 3W}
  \end{subfigure}\hfil
   \begin{subfigure}[t]{0.45\linewidth}
    \centering
    \includegraphics[width=0.9\linewidth]{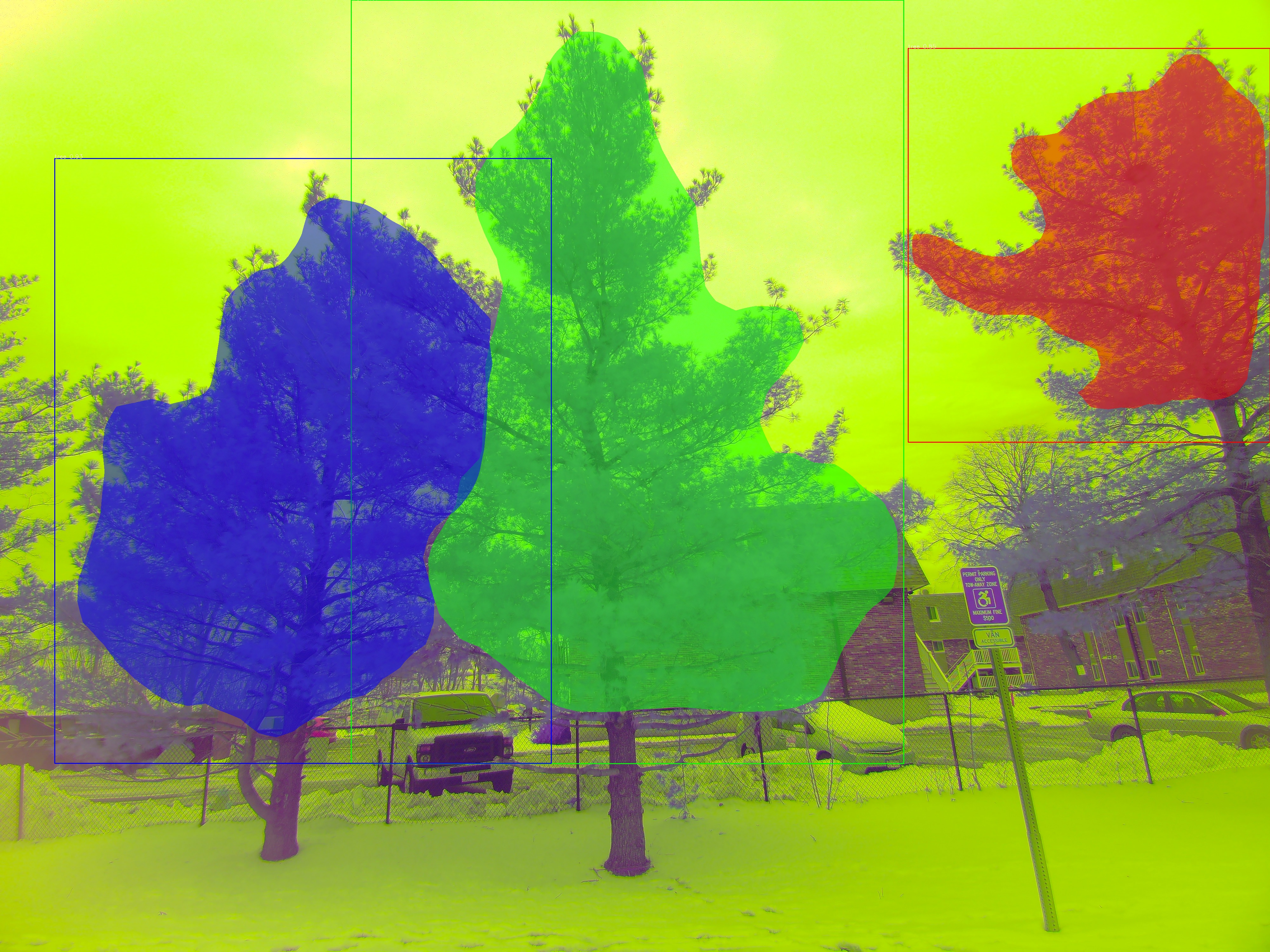}
    \caption{Segmentation output from our Custom Mask R-CNN instance segmentation model trained using transfer learning}
  \end{subfigure}
  \caption{Performance of our custom Mask R-CNN. Notice how the model detects each instance of the tree canopy in the image and considers all the other objects as background}
 \label{fig:maskrcnnOutput} 
\end{figure}

Our dataset consists of a relatively small number of images to train a deep learning model like Mask R-CNN from scratch. Transfer learning combined with data augmentation was employed in order to develop a custom model by using an existing model pre-trained on a different dataset. For this, we used a Mask R-CNN pre-trained \cite{matterport_maskrcnn_2017} on COCO \cite{coco} (a dataset with 330K images) with ResNet101 as the backbone. We re-trained only the head layers (the top layers without the backbone) on our dataset. The batch size was configured as 4 and the number of epochs was 10. The training was performed on the Google Cloud Platform with a N1 instance with 13GB memory and 2vCPUs. We also generated synthetic data by augmenting the original dataset with flips in the horizontal and vertical directions and applying Gaussian blur. This increased the training dataset size by 50\% and acted as a regularizer. The manually annotated dataset (refer Section \ref{sec:datasetCreation}) consisting of 51 images was split in the ratio of 70: 30 for training: testing. During re-training, each epoch took approximately 3 hours on the N1 instance. The training curve of the model is shown in Figure \ref{fig:trainingCurve}. It is seen from the training curve that only a small number of epochs are sufficient to reach the optimal validation loss on the test set owing to the retraining of only the head layers. The visual output results from our model are shown in Figure \ref{fig:maskrcnnOutput}.

\subsubsection{Model Quantization}
Mask R-CNN is a relatively heavy model both from training and inference points of view. Hence, the developed Mask R-CNN was optimized to run on the edge at the cost of possible minute performance reduction. For this, the model built on TensorFlow was converted to TensorFlow-lite with dynamic range quantization \cite{tensorflowQuantization}. Dynamic range quantization means that only the weights of the layers in 32-bit FLOAT of the full model are stored as 8-bit INTs while the activations of the layers are quantized during runtime. Our custom-made Mask R-CNN built over TensorFlow took around 15 seconds per inference of an image on a Raspberry Pi 3 while the TensorFlow-lite model reduced the inference time to 7 seconds with one-fourth of the CPU usage as the original TensorFlow model.

\section{Evaluation}
\label{sec:evaluation}
The system was evaluated using a dataset obtained from the municipality of Cambridge, USA as ground truth reference. We also conducted three data collection experiments to collect data of urban trees using the GreenScan system. In this section, we elaborate on this dataset and the data collection experiments followed by the obtained results.

\subsection{(Ground truth) Tree Health Dataset}
\label{sec:groundTruthDatabase}
Municipalities in cities obtain tree health data through city-wide surveys over years. For instance, in the city of Cambridge, USA, a survey is performed every 5 years whereas, for the city of Delft, The Netherlands, a survey is performed every 1 year to 2 years depending on the previously rated condition of the tree.
For the evaluation of our work, we obtained the tree health dataset for the city of Cambridge, USA through the Cambridge Urban Forest Master Plan \cite{CUFMPAppendix} to be used as ground truth reference. A 2018 dataset was obtained from the municipality. This dataset was created through a combination of manual in-person arborist visits, satellite-based remote sensing and aerial LiDAR \cite{CUFMPAppendix}. The dataset classifies the health conditions of trees into three categories, namely good, poor, and fair. The dataset contains information about 47,063 trees out of which 35,821 are in good health, 5176 are in fair health and 6066 are in poor health. Hence, most of the trees ($ > 75 \%$) are rated as having a good health condition. In addition, the dataset contains information about the tree species, common name, the satellite-based NDVI, the latitude and the longitude, location, the shape length and shape area of the canopy and other parameters. This dataset was provided as Shapefiles (.shp, a data format used by GIS (Geographical Information Systems)) and was loaded to the online platform CARTO \cite{carto} (a GIS and spatial analysis tool). On a side note, the staleness of data in terms of time also necessitates the advancement in this field of tree health monitoring.

\subsection{Data Collection Experiments}
\label{subsec:dataCollection}
\begin{figure}[ht]
    \centering
    \includegraphics[width=\linewidth]{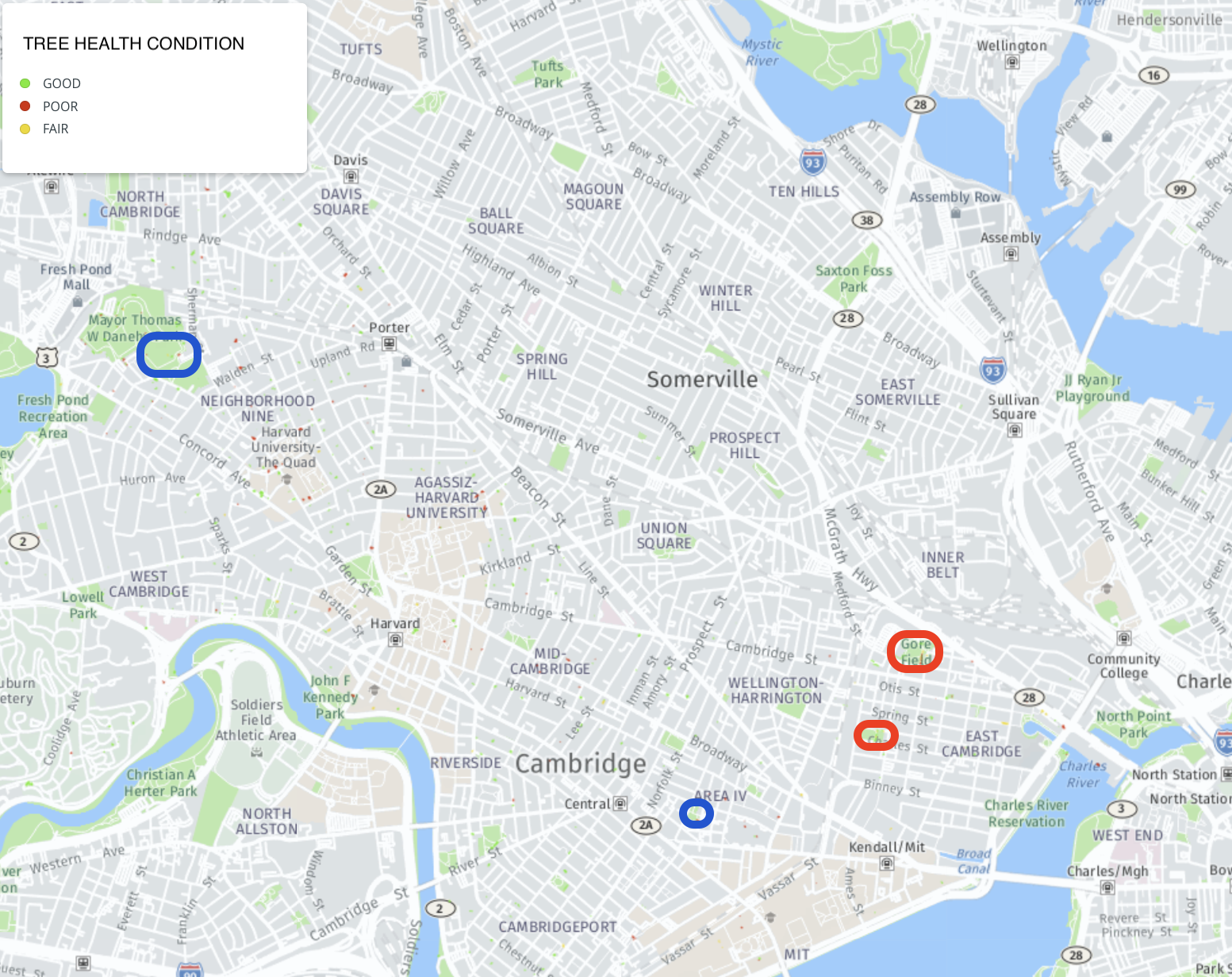}
    \caption{The trees were analysed in these locations. The red boxes indicate the Red Pine trees and the blue boxes indicate the Eastern White Pine trees.}
    \label{fig:dataCollectionSites}
\end{figure}

We collected multispectral (RGN) and thermal images through the developed system on three separate days in Cambridge, USA during the month of February 2022. A push button was used as the event trigger for the system. Hence, we employed the developed GreenScan system as a citizen science project with the 3D printed casing (pedestrian moving at walking speed in a straight line at a distance of 8m - 20m from the tree). In total, we collected data for 49 trees spread over two species namely Red Pine and Eastern White Pine trees. The multispectral imaging sensor was configured with a shutter speed of 1/60s and ISO of 50. The thermal imaging sensor was configured to measure temperature in range of (-10, 40) $^\circ $C. The sites of data collection experiments are shown in Figure \ref{fig:dataCollectionSites}, chosen based on the species and accessibility.

\textbf{Species Constraints:}
There are two types of trees namely evergreen and deciduous trees. During winters, deciduous trees lose their leaves, thus hampering NDVI calculation. Hence, our analysis was constrained to evergreen trees due to data collection in the winter. The species namely Red pine and Eastern White Pine were selected because they are evergreen and they are the most widespread and easily accessible evergreen trees found from CARTO in the city of Cambridge.

\textbf{Data Cleaning:}
During the first day of the data collection experiments, the Raspberry Pi hung up due to unknown reasons leading to a forced restart. On the third day of the experiments, owing to cold temperatures, the power supply had to be changed during data collection. These interruptions and restarts resulted in unstable values for a sequence of readings related to the canopy temperature by the thermal imaging sensor. As a result, these 11 data points were removed from our dataset generated using data collection experiments. In the end, our dataset was reduced to contain 40 trees. Distribution of the data collected from each of the tree species after data cleaning is shown in Table \ref{tab:dataCollection}.

\begin{table}[ht]
\caption{Distribution of trees after data cleaning}
\begin{tabular}{|l|l|l|}
\hline
 \textbf{Species}                  & \textbf{Number of Trees} & \textbf{Health Distribution}  \\ \hline
Red Pine           & 26                       & Good: 15 Fair: 7 Poor: 4                \\ \hline
Eastern White Pine & 14                        & Good: 5 Fair: 1 Poor: 8                 \\ \hline
\end{tabular}
\label{tab:dataCollection}
\end{table}

\subsection{Performance of custom Mask R-CNN}

To measure the performance of our custom Mask R-CNN model, we calculated the standard evaluation metrics as used by COCO \cite{cocoEval}. Specifically, we measured $mAP$ (mean Average Precision) / $AP$ (Average Precision) at different IoU (Intersection over Union) thresholds (as per \cite{cocoEval}). The performance of our custom Mask R-CNN with and without quantization is shown in Table \ref{tab:performaceMRCNN}. A comparison of inference time and model size comparing both the full model and the quantized model is also shown in Table \ref{tab:performaceMRCNN}. In order to measure the stability of our results, k-Fold cross-validation \cite{kfold} was also performed with k=3, to evaluate the performance of the model on different training and test splits as shown in Table \ref{tab:crossValidationResults}. These results for different train-test splits as shown in Table \ref{tab:crossValidationResults} showcase the reliability of our model. 

\begin{figure}[ht] 
\centering
  \begin{subfigure}[t]{0.45\linewidth}
    \centering
    \includegraphics[width=0.9\textwidth]{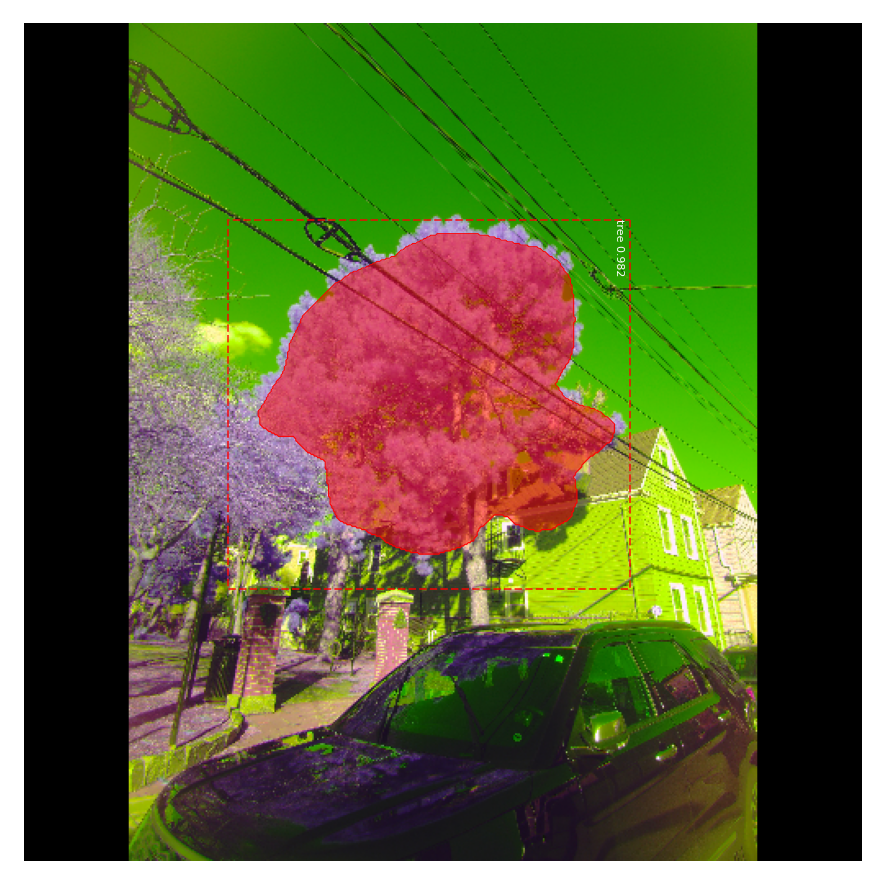} 
    \caption{Segmentation output from Mask R-CNN using full Tensorflow model}
  \end{subfigure}\hfil
   \begin{subfigure}[t]{0.45\linewidth}
    \centering
    \includegraphics[width=0.9\linewidth]{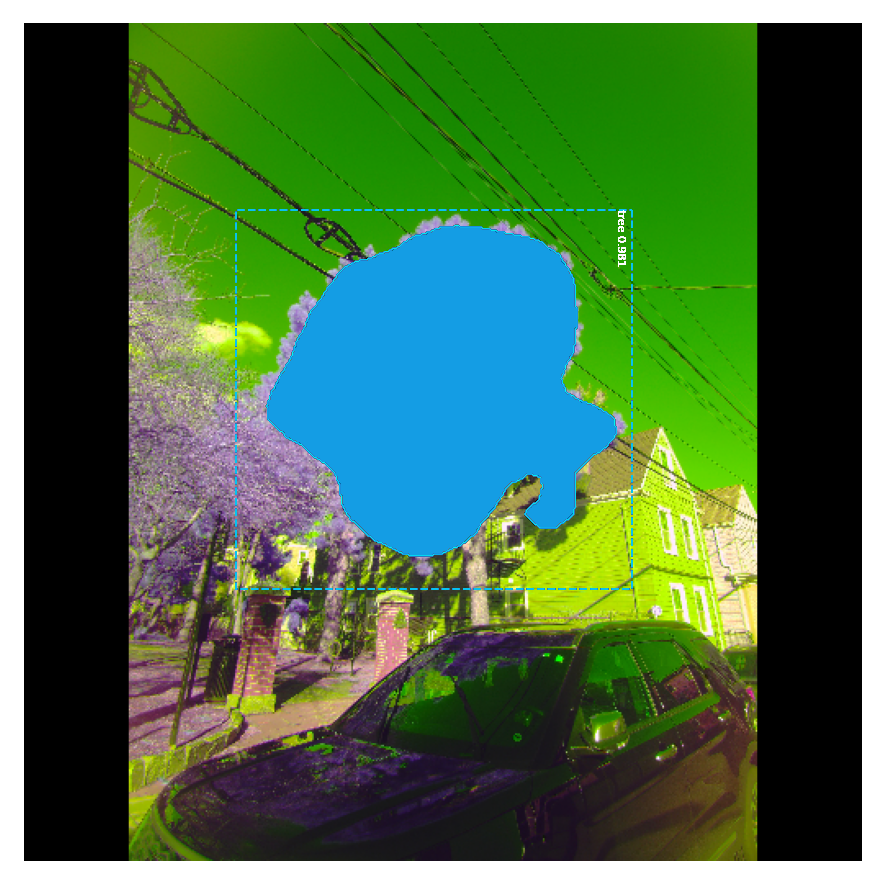}
    \caption{Segmentation output from Mask R-CNN using Tensorflow-lite (quantized)}
  \end{subfigure}
  \caption{Outputs from the full and quantized custom Mask R-CNNs}
 \label{fig:maskrcnnOutputComparison} 
\end{figure}

From Table \ref{tab:performaceMRCNN}, it is seen that there is no significant reduction in performance using quantization. The inference time of the quantized model is half compared to the non-quantized model along with the reduction of the model size. 
An example of segmentation outputs generated by the full model and quantized model on the same image is also shown in Figure \ref{fig:maskrcnnOutputComparison}. This also showcases the similar performance for both the full and the optimised model in a visual form.

\begin{figure}[ht]
    \centering
    \includegraphics[width=0.9\linewidth]{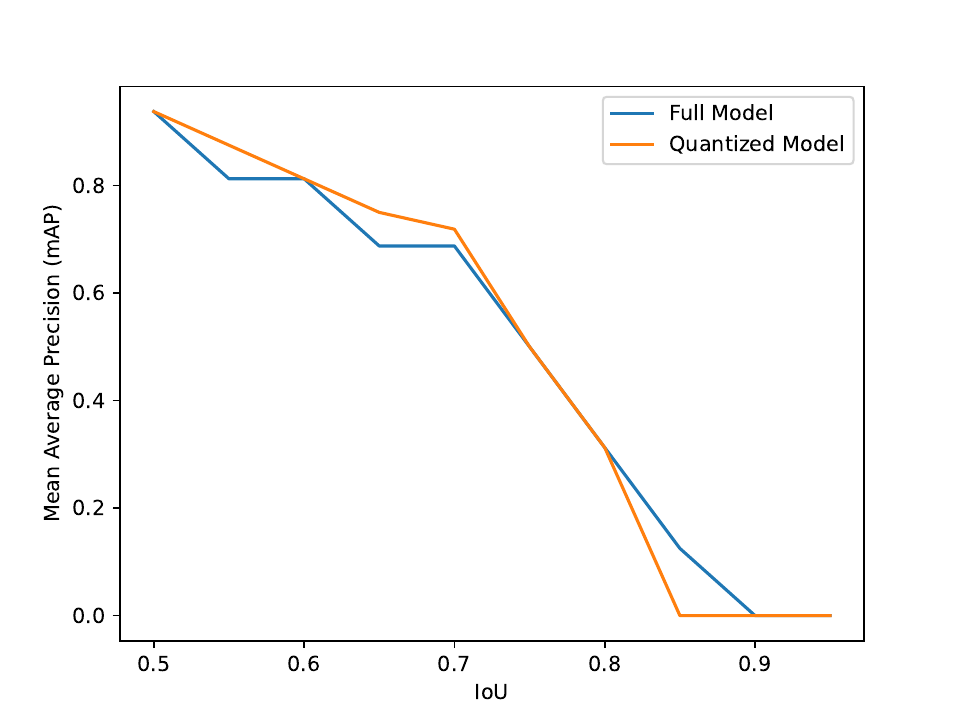}
    \caption{The AP scores with increasing IoU thresholds as per COCO metrics \cite{coco} for the full and quantized models}
    \label{fig:mapScores}
\end{figure}

From Table \ref{tab:performaceMRCNN}, it may appear that the $AP^{(IoU=0.5:0.95:0.05)}$ \cite{cocoEval} for the quantized model is increased slightly compared to the full model. On further exploring this anomaly, it was found that this behaviour is exhibited  due to our annotated dataset where most images contain only one full tree canopy as ground truth. Thus, a model (non-quantized model) generalising better to find partially visible tree canopies, in addition to the full tree canopy is penalised in terms of Precision (False Positive). Further, it is seen from Figure \ref{fig:mapScores} that the performance of the quantized model decreases more than the full model at higher IoUs (IoU= 0.85 for quantized model compared to 0.90 for full model) signifying that it is slightly poorer at object localisation compared to the full model.

\begin{table*}[ht]
\centering
\caption{Performance of custom R-CNN model (Full and Quantized model)}
\begin{tabularx}{\linewidth}{|X|l|X|X|l|l|}
\hline
\textbf{Model}                                    & \textbf{$AP^{(IoU=0.5:0.95:0.05)}$} & \textbf{$AP^{(IoU=0.5)}$} & \textbf{$AP^{(IoU=0.75)}$} & \textbf{$Inference Time$} & \textbf{$Model Size$}  \\ \hline
\textbf{Custom Mask R-CNN  TF}                      &   0.489         &  0.938           &  0.500    & 15s   & 255.9 MB      \\ \hline
\textbf{Custom Mask R-CNN  TF-lite (Dynamic Quantization)} &  0.491          &  0.938           & 0.500  & 7s   & 65 MB            \\ \hline
\end{tabularx}
\label{tab:performaceMRCNN}
\end{table*}

\begin{table}[h]
\centering
\caption{Results of 3-Fold Cross Validation of custom Mask R-CNN model}
\begin{tabularx}{0.8\linewidth}{|l|X|X|X|}
\hline
\textbf{Cross Validation Fold} & \textbf{1} & \textbf{2} & \textbf{3} \\ \hline
\textbf{$AP^{(IoU=0.5)}$}                    & 0.82      & 0.87      & 0.75        \\ \hline
\end{tabularx}
\label{tab:crossValidationResults}
\end{table}

\subsection{Results for the health of trees}
We extracted three parameters from the ground truth dataset namely Ground Truth Condition (Health), Remote NDVI, and Area of the tree (measured using aerial LiDAR) from all the parameters present in the dataset.

\begin{figure}[h]
    \centering
    \includegraphics[width=0.9\linewidth]{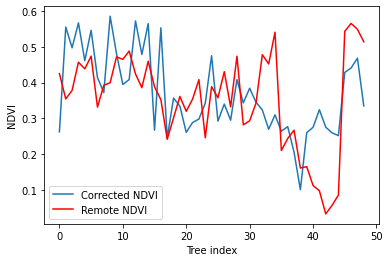}
    \caption{Variation of measured NDVI vs Remote NDVI for trees observed during data collection experiments. The tree index refers to an individual tree ID.}
    \label{fig:correctedNDVivsRemote}
\end{figure}

A comparison of our system-measured NDVI and Remote NDVI is shown in Figure \ref{fig:correctedNDVivsRemote}. As seen in this Figure, our measured NDVI is distributed similarly to the Remote NDVI for an individual tree (denoted by Tree Index in Figure \ref{fig:correctedNDVivsRemote}) .

\begin{figure}[h]
    \centering
    \includegraphics[width=0.7\linewidth]{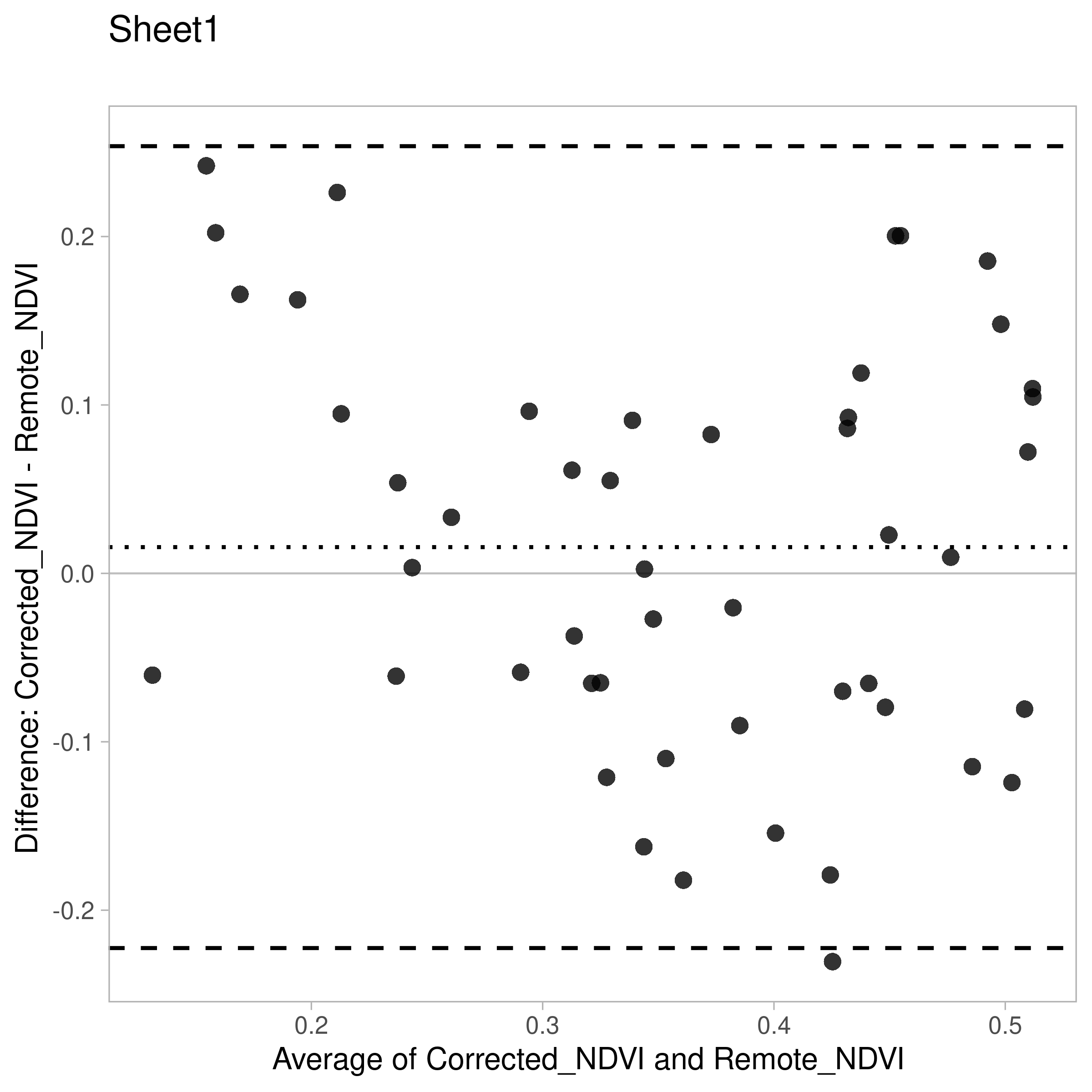}
    \caption{The Bland-Altman plot showcasing the agreement between the our measured NDVI and Remote NDVI. The dashed-middle line shows the mean difference. The top most and bottom most lines refer to the 95\% limits of agreement respectively.}
    \label{fig:baplot}
\end{figure}

Two datasets can be highly correlated but strongly disagree. Hence, a Bland-Altman plot \cite{blandAltmanTest} widely used to showcase the agreement between two monitoring methods measuring the same attribute was utilised. It plots the difference between the corresponding measurement values against the average of those values. From the Bland-Altman plot shown in Figure \ref{fig:baplot}, it is seen that there is strong agreement between the two methods (Remote NDVI and our measured NDVI) with all points (representing data for each tree) except one lying within the 95\% limits of agreement (average difference within ± 1.96 standard deviation of the difference, as the difference between values follows a normal distribution).

\begin{figure}[ht]
    \centering
    \includegraphics[width=0.9\linewidth]{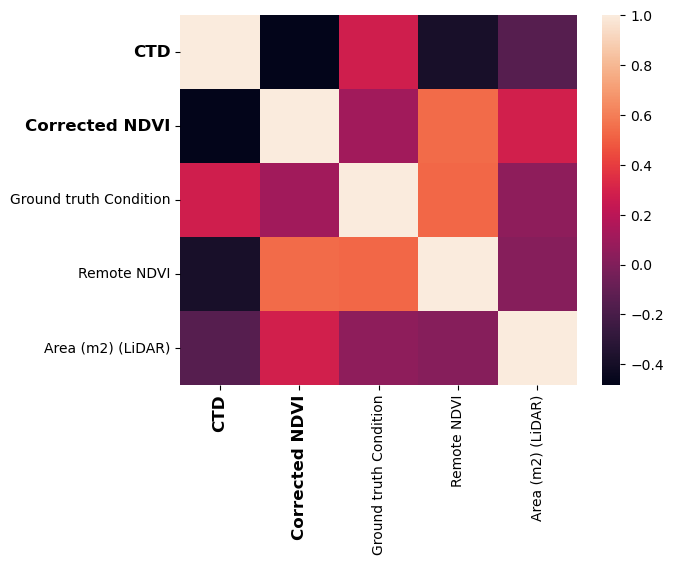}
    \caption{Correlation matrix between our measured values (in bold) and parameters from ground truth dataset}
    \label{fig:correlationMatrix}
\end{figure}

Pearson's correlation coefficient (r) was also measured to calculate the strength of the linear relationship between our measured values and ground truth data. The correlation matrix comprising all of our measured values with the three ground truth parameters namely Ground Truth Condition, Remote NDVI, and Area is shown in Figure \ref{fig:correlationMatrix}. Further, the correlation results between the measured NDVI and CTD with ground truth parameters is shown in Table \ref{tab:correlationResults}. 


\begin{table}[h]
\centering
\caption{The correlation between our measured values and ground truth parameters}
\begin{tabular}{|cc|c|c|}
\hline
\multicolumn{2}{|c|}{\textbf{Variables}}                                                    & \textbf{Pearson Correlation} & \textbf{Significant at} \\ \hline
\multicolumn{1}{|c|}{\textbf{Measured}} & \textbf{Ground Truth} & \textbf{(r)}  & \textbf{(p \textless 0.05)} \\ \hline
\multicolumn{1}{|c|}{NDVI} & Remote NDVI                                                    & 0.54                                                                       & Yes  \\ \hline
\multicolumn{1}{|c|}{CTD}  &  Health Condition & 0.28         & Yes                                                                                   \\ \hline
\multicolumn{1}{|c|}{NDVI} & \begin{tabular}[c]{@{}c@{}}Area (m2) \\ (LiDAR)\end{tabular}   & 0.28                                                                       & Yes \\ \hline
\multicolumn{1}{|c|}{CTD}  & \begin{tabular}[c]{@{}c@{}}Area (m2) \\ (LiDAR)\end{tabular}   & -0.15                                                                      & No  \\ \hline
\end{tabular}
\label{tab:correlationResults}
\end{table}

\begin{figure}[ht]
\centering
\begin{subfigure}[t]{\linewidth}
  \centering
    \includegraphics[width=\linewidth]{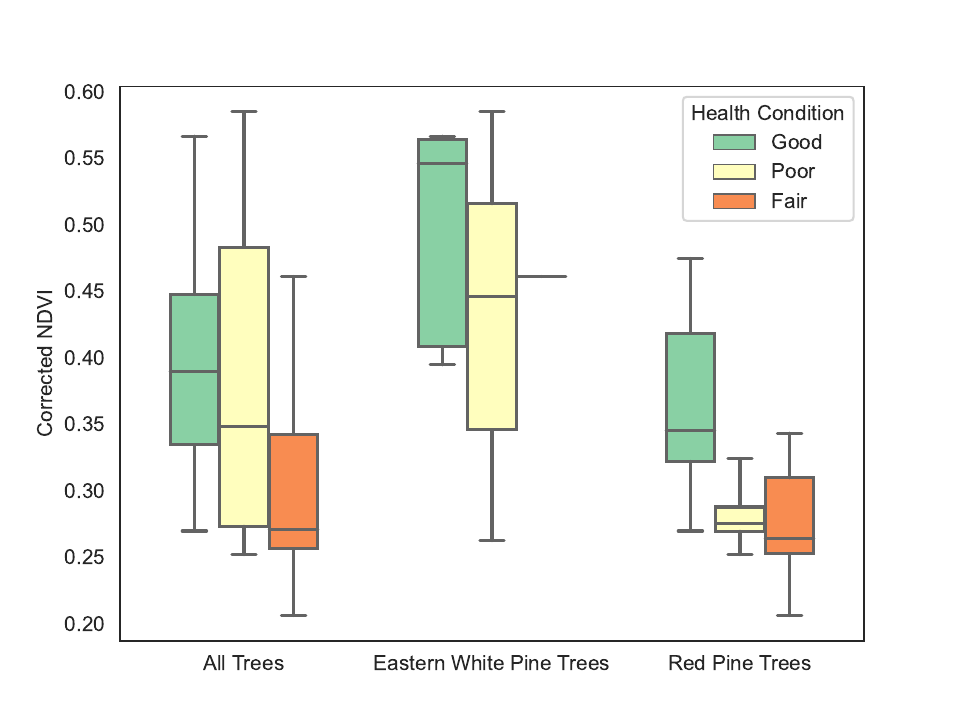}
    \caption{ The distribution of NDVI for the trees}
    \label{fig:ndviAllTrees}
\end{subfigure} \hfil
\begin{subfigure}[t]{\linewidth}
  \centering
    \includegraphics[width=\linewidth]{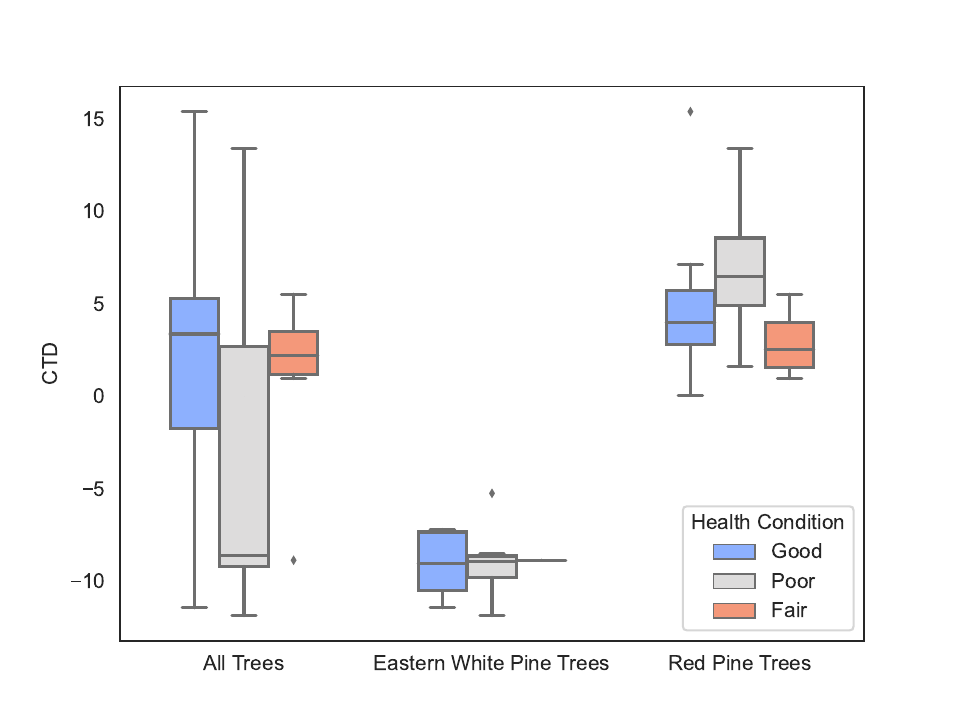}
    \caption{ The distribution of CTD for the trees }
    \label{fig:ctdAllTrees}
    \end{subfigure}
 \caption{ The distribution of NDVI and CTD for the trees with respect to health }
    \label{fig:distributionNDVICTD}
\end{figure}

The distribution of CTD and NDVI with respect to ground truth health conditions is shown in Figure \ref{fig:distributionNDVICTD}. From the NDVI distribution in Figure \ref{fig:ndviAllTrees}, it is seen that the extent of agreement of NDVI with respect to the ground truth health conditions varies with respect to the species. For red-pine trees, the trees in good health condition have higher measured NDVI values than trees in poor and fair condition. A similar conclusion is drawn from the CTD distribution in Figure \ref{fig:ctdAllTrees}. The mean NDVI and CTD for each species is also shown in Table \ref{tab:ndviCtdSpecies}.

\begin{table}[ht]
\centering
\caption{The mean of measured NDVI and CTD across species and health}
\begin{tabularx}{\linewidth}{|X|l|l|l|}
\hline
Species / Health           & \textbf{Good} & \textbf{Fair} & \textbf{Poor} \\ \hline
\textbf{Red Pine (NDVI)}           & $0.37 \pm 0.07$   & $0.28 \pm 0.05$    & $0.28 \pm 0.03$     \\ \hline
\textbf{Eastern White Pine (NDVI)} & $0.49 \pm 0.08$   & 0.46          & $0.43 \pm 0.12$   \\ \hline
\textbf{Red Pine (CTD)}           & $4.63 \pm 3.64$   & $2.89 \pm 1.78$    & $6.99 \pm 4.85$     \\ \hline
\textbf{Eastern White Pine (CTD)} & $-9.1 \pm 1.88$   & -8.59          & $-9.1 \pm 1.88$   \\ \hline
\end{tabularx}
\label{tab:ndviCtdSpecies}
\end{table}

\subsubsection{High-level tree health analysis}
From Figure \ref{fig:correlationMatrix}, it is clear that there is almost no correlation between NDVI and CTD. Thus, they are independently measuring two different attributes related to tree health and useful to incorporate in the system. In recent works such as \cite{correlationSatellites}, the correlation between remote NDVI measured using two different satellites was found to be 0.74 (moderately strong). In this work, the Table \ref{tab:correlationResults} shows moderately strong correlation (r=0.54 with $p<0.05$) between our measured NDVI (ground based) and remote NDVI. This moderately strong correlation serves to showcase the validity of our approach for ground-based NDVI measurement using multispectral imaging sensors. 

Since there is no ground truth reference attribute for CTD which indicates water stress of trees, we checked the correlation of CTD with ground truth health condition as shown in Table \ref{tab:correlationResults}. The weak-moderate correlation (r=0.28 with $p<0.05$) between CTD and ground truth tree health condition can be attributed to the skewed distribution of the dataset where more trees are rated as having good conditions compared to poor and fair conditions. Further analysis of CTD distribution for all trees in Figure \ref{fig:ctdAllTrees} shows the high variability of CTD for trees in poor condition leading to this overall weak-moderate correlation.

\subsubsection{Species-wise tree health analysis}

From the NDVI distributions for Eastern White pine in Figure \ref{fig:ndviAllTrees}, it is seen that the good condition trees are generally distributed to have higher NDVI values than poor and fair condition trees. Thus, a simple threshold based classification algorithm can easily flag trees which might not specifically be in good health conditions. At a scale of tens of thousands of trees in a city, this can lead to a significant amount of cost savings.

From Figure \ref{fig:ctdAllTrees}, while a higher CTD is found for red pine trees in poor condition than good and fair health condition trees, the same pattern is not applicable for eastern white pine trees. This inference about CTD is similar to earlier works such as \cite{thermalWaterStress} \cite{Ballester2013UsefulnessOT}, where the tree species under observation has a significant influence on the results obtained from thermal imaging. Hence, further studies with varied species are required to measure the stability of CTD with respect to ground truth health conditions.

\section{Limitations and future work}
Subsequent investigations stemming from this work and using our approach as a foundational framework are expected to illuminate and solve several novel challenges, subset of which are delineated as follows.

\subsubsection{Feasibility of modelling-based classification}
\begin{figure}[ht]
    \centering
    \includegraphics[width=0.9\linewidth]{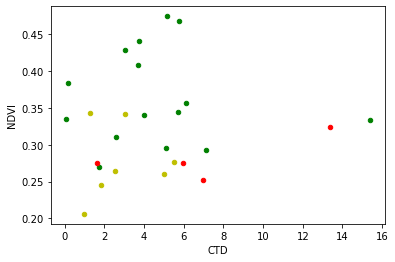}
    \caption{Scatter plot between NDVI and CTD (in $^\circ $C) for red pine trees. The color of the points indicates the ground truth health with red denoting poor, yellow denoting fair, and green denoting good condition trees.}
    \label{fig:NDVIvsCTDPlot}
\end{figure}

From the correlation matrix in Figure \ref{fig:correlationMatrix}, it is seen that is no correlation between CTD and NDVI values. Hence, in autonomous models to classify tree health, both these measured parameters are useful features. A scatter plot between NDVI and CTD values for red pine trees in shown in Figure \ref{fig:NDVIvsCTDPlot}. From the scatter plot, it is seen that most of the fair and poor-condition trees are concentrated around a cluster between NDVI (0.20-0.35) and CTD (0-7). Hence, simple white-box machine learning algorithms like SVMs (support vector machines) with kernel \cite{svm} or logistic regression classifiers \cite{lreg} can autonomously distinguish between good, and poor or fair condition trees. Further, the methodology can be expanded by adding human-in-the-loop-validation at intermediate steps to enhance the performance of the system.

\subsubsection{Direction of movement and Robust positioning system}
At present, our methodology is contingent upon aligning the system consisting the imaging sensors with the trees' orientation. However, given the intended practical application in real-world scenarios, the angle between the tree canopy and the camera's direction could potentially impact the segmentation of tree canopies. This can be improved by adding a simple image selection algorithm such that the majority of image frame is occupied by pixels belonging to a tree canopy. Furthermore, our current positioning method relies on GPS coordinates sourced from the tree survey dataset and the GNSS module within the system, which inherently faces uncertainties in positioning. Thus, an alternative positioning approach utilising RTK (Real-Time Kinematics) can enhance the positioning robustness.

\subsubsection{Scalability in different weather conditions and geographical boundaries}
The effect of different weather conditions with reduction of visibility, and sunlight directly facing the imaging sensor lenses needs further exploration. Secondly, the deployment and validation of the system in cities with different topographies and with geographical domain shifts can help in enhancing the generalization of the approach using large-scale training and validation datasets.

\section{Conclusion}
Urban greenery provides various environmental services such as carbon sequestration, and cooling making them essential for building climate-adaptive cities. Currently, urban trees are experiencing atypical amounts of natural and human-induced stresses leading to their volatile health.  Yet, high costs make it infeasible for cities to perform frequent inspections on a large scale, leading to adverse health conditions being discovered only after severe damage. The current popular methods for monitoring the health of urban trees rely on an in-person inspection performed by arborists and remote sensing based on satellites or airborne imagery. However, all these methods are riddled with various challenges involving scalability, spatio-temporal resolutions, and quality of assessment. In this work, we developed a novel system called GreenScan to measure tree health autonomously from the ground level in urban cities. The GreenScan system fuses data from low-cost thermal and multispectral imaging sensors using custom computer vision models optimised for efficiency to generate the tree health indexes namely NDVI and CTD. The custom Mask R-CNN model fine-tuned using transfer learning was employed to fuse the data collected by the imaging sensors on the edge device. Deployment can be performed both in a drive-by sensing paradigm on moving vehicles such as taxis and garbage trucks or in a citizen-science sensing paradigm by humans. Initial evaluation of the system was performed through data collection experiments in Cambridge, USA. The custom Mask R-CNN developed performed admirably, with an $AP^{IoU=0.50} = 0.938$ despite the small dataset used for training. The tree health analysis revealed moderately-strong correlation between our measured NDVI and the remote NDVI obtained from the ground truth dataset. Further, our measured NDVI distributions can be used to flag trees that are specifically not in good health conditions. For the measured CTD, a pattern with a theoretical agreement was applicable for one of the species observed. However, further large-scale evaluation studies over multiple species would help in improving the generalisability of the system. In essence, this work illustrates the potential of autonomous ground-based urban tree health monitoring on city-wide scales at high temporal resolutions and motivates future research at the intersection of environmental science and computer science.


\bibliographystyle{IEEEtran}
\bibliography{references}

\end{document}